\documentclass{ckm}                 % twocolumn proceedings style
\usepackage{txfonts}            % if you have it, to get Times family
\confname{Workshop on the CKM Unitarity Triangle, IPPP Durham, April
  2003}

\title{Charm Phenomenology for CKM Parameters}

\author{Alexey A. Petrov}

\address{Department of Physics and Astronomy, Wayne State University, Detroit, MI 48201, USA}

\def\barD{\overline D{}^0}

\def\DDbar{D^0-\overline D{}^0}
\def\BBbar{B^0-\overline B{}^0}

\def\D0bar{\overline D{}^0}
\def\K0bar{\overline K{}^0}

\def\3bar{\overline{3}}

\def\15bar{\overline{15}}
\def\24bar{\overline{24}}
\def\42bar{\overline{42}}
\def\60bar{\overline{60}}

\def\cal{{\it}}

\def\beq{\begin{equation}}
\def\eeq{\end{equation}}
\def\beqa{\begin{eqnarray}}
\def\eeqa{\end{eqnarray}}
\def\bea{\begin{eqnarray}}
\def\eea{\end{eqnarray}}

\begin{document}

\begin{abstract}
We review several key aspects of charm phenomenology which are important for 
the general program of determination of matrix elements of the 
Cabbibo-Kobayashi-Maskava (CKM) matrix.
We argue that charm physics plays an important role in reducing uncertainties 
in non-perturbative QCD parameters whose values are required for understanding 
of underlying quark-level processes. We also review the current status of
searches for new physics in mixing and  $CP$ violation 
in charmed mesons both at the currently operating and proposed facilities.
\end{abstract}

%% \maketitle needs to be after the author and address info and the
%% abstract... 
\maketitle

%% standard LaTeX from here on...

%%%%%%%%%%%%%%%%%%%%%%%%%%%%%%%%%%%%%%%%%%%
\section{Introduction}

Charm physics plays a unique dual role in the modern investigation of flavor physics. 
Charm decay and production experiments provide valuable checks and supporting measurements 
for studies of CP-violation in measurements of CKM parameters in b-physics, as 
well as outstanding opportunities for searches for new physics.

Historically, many methods of heavy quark physics have been first tested in charmed hadrons.
The fact that a b-quark mainly decays into a charm quark makes charm physics an integral
part of any b-physics program. As we shall see in many cases, direct measurements of 
charm decay parameters directly affect the studies of fundamental physics in 
B decays.

Any search for new physics falls into two distinct categories: {\it direct}, when new
physics particles are directly produced in the experiment and {\it indirect}, when
they only affect low energy observables via quantum corrections. In turn,
indirect searches can also be roughly divided into three classes, depending on
which observables or processes are used in the search: 1) Processes forbidden in 
the Standard Model (SM), i.e. absent at any order in expansion in any of the 
coupling constants of SM. An example of those could be provided by charge-violating 
transitions $B^+ \to \mu^+ \mu^-$ or $D^0 \to \mu^+ \mu^+$.
2) Processes forbidden in the Standard Model at tree level, but possible at higher
orders. Examples of those processes involve flavor-changing neutral current (FCNC)
transitions, such as familiar $B \to X_{d,s} \gamma$, $D \to X_u \gamma$ or 
$\BBbar$/$\DDbar$ mixing. 3) Processes allowed in the Standard Model. 
Examples of the tests that belong to this category include examinations of the relations 
between different observables that are satisfied in the Standard Model, 
but not necessary in general. An example of those are the triangle relations 
among CKM parameters extracted form several different decays. 

The Standard Model is a very constrained system, which implements a remarkably simple
and economic description of all CP-violating processes in the flavor sector by 
a single CP-violating parameter, the phase of the CKM matrix. This fact relates 
all CP-violating observables in bottom, charm and strange systems and provides an 
excellent opportunity for searches of physics beyond the Standard Model.

%%%%%%%%%%%%%%%%%%%%%%%%%%%%%%%%%%%%%%%%%%%
\section{Supporting B-physics measurements}

One of the major goals of the contemporary experimental $B$ physics program is an accurate 
determination of the CKM parameters. As we shall see below, inputs from charm decays are important 
ingredient in this program, both in the extraction of the angles and sides 
of the CKM unitarity triangle.

%%%%%%%%%%%%%%%%%%%%%%%%%%%%%%
\subsection{Extraction of the CKM angle $\gamma$}

Some of the cleanest methods of the determination of the CKM phase 
$\gamma=\arg\left[-V_{ud}V_{ub}^*/V_{cd}V_{cb}^*\right]$ involve
the interference of the $b \to c\bar{u}s$ and $b \to u\bar{c}s$ quark-level
transitions~\cite{Fleischer:2003yc,Gershon:2003zk}. A way to arrange for an interference 
of those seemingly different 
processes was first pointed out in~\cite{Gronau:1991dp}. It involves interference
of two {\it hadronic} decays $B^+ \to D^0 K^+ \to f K^+$ and 
$B^+ \to \barD K^+ \to f K^+$, with $f$ being any common final state for $D^0$ 
and $\barD$ decays. Since then, many different methods have been proposed, mainly differing by the 
$f K$ final state and paths of reaching it: a combination of the Cabibbo-favored (CF) and 
doubly-Cabibbo suppressed decays (DCSD)
$D^0(\barD) \to K^+ \pi^-$~\cite{Atwood:1996ci}, 
singly-Cabibbo suppressed decays (SCSD)
$D^0(\barD) \to K K^*$~\cite{Grossman:2002aq}, Cabibbo-favored decays employing
large $K^0-\overline{K^0}$ mixing 
$D^0(\barD) \to K_S \pi^0$~\cite{Fleischer:2003aj}, as well as multibody versions of these
transitions~\cite{Giri:2003ty}. For similar methods involving interference of the 
{\it initial} state, see~\cite{Falk:2000ga}.

Each of those methods involve observations of several $B$ decay modes,
so all of the hadronic unknowns, such as the ratio 
\beq\label{BDratio}
A(B^+ \to D^0 K^+)/A(B^+ \to \barD K^+)=r_B e^{i \left(\gamma+\Delta_B\right)},
\eeq
or $A(D^0 \to f)/A(\barD \to f)$, as well as $\gamma$, can be determined from 
the $B$-decay measurements only.
Yet, all of those methods will benefit from the separately performed charm decay 
measurements, the primary reason being the need for large statistics in 
any experimental determination of $\gamma$. 
For example, an ADS method~\cite{Atwood:1996ci} employs a ratio of a DCS to CF decays 
$A(D^0 \to K^+ \pi^-)/A(\barD \to K^+ \pi^-)=r_D e^{i \delta_D}$.
The accuracy of this method will be greatly improved if $r_D$ and $\delta_D$ are measured 
separately. While the separate measurement of $r_D$ is already available,
determination of $\delta_D$ will became possible at the upcoming tau-charm factories (TCF) 
at Cornell (CLEO-c) and Beijing (BES-III).

The basic idea is to make use of an important experimental advantage of TCF: 
operation at the $D^0 \barD$ threshold, where the 
$D^0$ and $\barD$ are produced in the quantum-mechanically coherent initial 
state~\cite{Silva:1999bd,GGR}. Since the initial $D^0 \barD$ state is prepared as 
\begin{eqnarray} \label{TCFinitial}
|D \barD \rangle_L = \frac{1}{\sqrt{2}} 
\left \{
| D^0 (k_1)\barD (k_2) \rangle +
(-1)^L | D^0 (k_2)\barD (k_1) \rangle
\right \},
\end{eqnarray}
where $L$ is the relative angular momentum of two $D$ mesons, 
$CP$ properties of the final states produced in
the decay of $\psi(3770)$ are anti-correlated,
one $D$ state decayed into the final state with
definite $CP$ properties immediately identifies or tags
$CP$ properties of the state ``on the other side.'' 
That is to say, if one state decayed into, say $\pi^0 K_S$ 
with $CP=-1$, the other state is ``CP-tagged'' as being in the 
$CP = +1$ state. This allows one to measure $\cos \delta_D$.
In order to see this, let us write a triangle relation,
\begin{equation}
\sqrt{2} A(D_{\pm} \to K^- \pi^+) = A(D^0 \to K^- \pi^+) \pm 
A(\barD \to K^- \pi^+),
\end{equation}
which follows from the fact that, in the absence of CP-violation in charm,
mass eigenstates of the neutral $D$ meson coincide with its CP-eigenstates,
\beq \label{CPeig}
\sqrt{2} | D_\pm \rangle = | D^0 \rangle \pm | \barD \rangle.
\eeq
This implies a relation for the branching ratios,
\begin{equation} \label{cos1}
1 \pm 2 \cos \delta_D \sqrt{r_D} = 
2 \frac{Br(D_{\pm} \to K^- \pi^+)}{Br(D^0 \to K^- \pi^+)}, 
\end{equation}
where we used the fact that $r_D \ll \sqrt{r_D}$ and neglected 
$CP$ violation in mixing, which could undermine the
CP-tagging procedure by splitting the CP-tagged state
on one side into a linear combination of CP-even and
CP-odd states. Its effect, however, is completely negligible here.
Now, if both decays of $D_+$ and $D_-$ are measured, 
$\cos \delta_D$ can be obtained from the asymmetry
\begin{equation} \label{cos2}
\cos \delta_D = 
\frac{Br(D_{+} \to K^- \pi^+)-Br(D_{-} \to K^- \pi^+)}
{2 \sqrt{r_D} Br(D^0 \to K^- \pi^+)}. 
\end{equation}
Both Eqs.~(\ref{cos1}) and (\ref{cos2}) can be used to extract 
$\delta_D$ at TCF. Similar measurements are possible for other 
$D$ decays~\cite{Rosner:2003yk}. 

In addition, some methods require cross-checks in the charm sector that might 
restrict the accuracy of these methods. A good example is provided by the original
GW method~\cite{Gronau:1991dp}, which does not take into account the possibility of 
relatively large $\DDbar$ mixing~\cite{Silva:1999bd}. This method relies on the simple 
triangle amplitude relation 
$\sqrt{2} A(B^+ \to D_\pm K^+)= A(B^+ \to D^0 K^+) \pm A(B^+ \to \barD K^+)$,
again provided by Eq.~(\ref{CPeig}).
An amplitude $A(B^+ \to D_\pm K^+)$ is measured with $D$ decaying to a 
particular CP-eigenstate. Neglecting $\DDbar$ mixing, angle $\gamma$ can then be extracted, 
up to a discrete ambiguity, from the measurements of 
$B^\pm \to f_{CP} K^\pm$ and $B^\pm \to D^0,\barD K^\pm$. In particular,
\beq
\Gamma[B^\pm \to f_{CP} K^\pm] \propto 1 + r_B^2 + 2 r_B c_\pm, 
\eeq
where $c_\pm=\cos (\gamma \pm \Delta_B)$ and $r_B$ and $\Delta_B$ are defined from 
the ratio Eq.~(\ref{BDratio}). Then,
\beq
\sin^2 \gamma = \frac{1}{2} \left[
1-c_+ c_- \pm \sqrt{\left(1-c_+^2\right) \left(1-c_-^2\right)}
\right].
\eeq
It is easy to see that $\DDbar$ mixing, if not properly accounted for, can
affect the results of this analysis. Indeed, 
taking $\DDbar$ mixing into account results in the modification of 
the definitions of $c_\pm$,
\bea
c_\pm&\to&\cos (\gamma \pm \Delta_B) \mp \frac{x}{2 r_B} \sin 2\theta_D 
\nonumber \\
&-&
\frac{y}{2 r_D}
\left[
2 \eta_f r_D \cos\left(\gamma+2\theta_D\pm\Delta_B\right)+\cos 2\theta_D
\right],
\eea
where 
\begin{eqnarray} \label{definition}
x \equiv \frac{m_2-m_1}{\Gamma}, ~~
y \equiv \frac{\Gamma_2 - \Gamma_1}{2 \Gamma},
\end{eqnarray}
with $m_{1,2}$ and $\Gamma_{1,2}$ being the masses and widths of D-meson mass eigenstates
$D_{1,2}$, $\eta_f$ is a CP-parity of $f_{CP}$, and $\theta_D$ is a CP-violating phase 
of $\DDbar$ mixing.
It is easy to see that $y \sim 1\%$ can impact the determination of $\gamma$ from these modes.
Thus, separately constraining $\DDbar$ mixing parameters will be helpful. We shall discuss 
those later.

%%%%%%%%%%%%%%%%%%%%%%%%%%%%%%
\subsection{Cross-checks of unitarity relations}

There are several unitarity triangles that involve charm inputs~\cite{Bigi:1999hr}.
Since all CP-violating effects in the flavor sector of the SM are 
related to the single phase of the CKM matrix, all of the CKM unitarity
triangles, including the ones with most charm inputs, have the same area.
This could provides a non-trivial check of the Standard Model, if
measurements of all sides of these triangles are measured with sufficient accuracy.
Unfortunately, ``charm triangles'' are ``squashed'', with one side being 
much shorter then the other two. For example, in terms of the Wolfenstein 
parameter $\lambda=0.22$, the relation 
$V_{td}V_{cd}^*+V_{ts}V_{cs}^*+V_{tb}V_{cb}^*=0$ has one side ${\cal O}(\lambda^4)$
with the other two being ${\cal O}(\lambda^2)$, while the relations
$V_{ud}V_{us}^*+V_{cd}V_{cs}^*+V_{td}V_{ts}^*=0$ and
$V_{ud}V_{cd}^*+V_{us}V_{cs}^*+V_{ub}V_{cb}^*=0$
have one side ${\cal O}(\lambda^5)$ with the other two being ${\cal O}(\lambda)$.
Even though the second relation could provide an interesting information about the 
top quark CKM parameters and the last one can be checked using only the tree-level
transitions, it is unlikely that the required accuracy will be achieved in the 
near future. 

In addition, tests similar to the ``first row unitarity'',
$\left|V_{ud}\right|^2+\left|V_{us}\right|^2+\left|V_{ub}\right|^2=1$
(which at the moment fails at the level of $2.2~\sigma$~\cite{Abele:2002wc}) are possible.
For example, 
$\left|V_{cd}\right|^2+\left|V_{cs}\right|^2+\left|V_{cb}\right|^2=1$,
It could provide an interesting cross-check on the value of $V_{cb}$ extracted in
B-decays, if sufficient accuracy on the experimental measurement of
$V_{cd}$ and $V_{cs}$ is achieved. 

%%%%%%%%%%%%%%%%%%%%%%%%%%%%%%
\subsection{Form-factors and decay constants}

Since $m_b,~m_c \gg \Lambda_{QCD}$, both charm and bottom quarks can be regarded as 
heavy quarks.
Naturally, heavy quark symmetry relates observables in B and D transitions.
As an example, let us consider measurements in the charm sector 
affect determinations of the CKM matrix elements relevant to top quark in 
$\BBbar$ mixing.

A mass difference of mass eigenstates in $\BBbar$ system can be written as
\beq
\label{offdiag}
\Delta m_d =
\!\! {\cal C}\left[\alpha_s^{(5)}(\mu)\right]^{-6/23} 
\left[1+\frac{\alpha_s^{(5)}(\mu)}{4 \pi} J_5 \right] 
\langle\bar B_d^0|{\cal O}(\mu)|B_d^0\rangle, \nonumber
\eeq
where ${\cal C}=G_F^2M_W^2 
\left({V_{tb}}^{*}V_{td}\right)^2 \eta_B m_B S_0(x_t)/\left(4 \pi^2\right)$ (see 
Ref.~\cite{Buchalla:1995vs} for complete definitions of the parameters in this expression). 
The largest uncertainty of about 30\% in the theoretical calculation 
is introduced by the poorly known hadronic matrix element
$
{\cal A} = \langle\bar B^0|{\cal O}(\mu)|B^0\rangle
$.
Evaluation of this matrix element is a genuine non-perturbative task, which can
be approached with several different techniques. The simplest approach 
(``factorization'')
reduces the
matrix element ${\cal A}$ to the product of matrix elements measured in
leptonic $B$ decays
$
{\cal A}^{f} = (8/3)
$
$
\langle\bar B^0|\bar b_L\gamma_{\sigma}d_L|0\rangle
\langle 0|\bar b_L \gamma^{\sigma}d_L|B^0\rangle = (2/3) f_B^2 m_B^2
$, where we employed the definition of the decay constant $f_B$,
\beq
\langle 0|\bar b_L \gamma_\mu d_L|B^0({\bf p})\rangle = i p_\mu f_{B}/2.
\eeq
A deviation from the factorization ansatz is usually described by the parameter
$B_{B_d}$ defined as
$
{\cal A} = B_{B_d} {\cal A}^{f}
$;
in factorization $B_{B_d}=1$. Similar considerations lead to an introduction of 
the parameter $B_{B_s}$ defined for mixing of $B_s$ mesons. It is important to
note that the parameters $B_{B_q}$ depend on the chosen renormalization scale 
and scheme. It is convenient to introduce renormalization-group invariant
parameters $\hat B_{B_q}$
\beq
\label{Bhat}
\hat B_{B_q} =
\!\! \left[\alpha_s^{(5)}(\mu)\right]^{-6/23}
\left[1+\frac{\alpha_s^{(5)}(\mu)}{4 \pi} J_5 \right]
B_{B_q}.
\eeq
We provide averages of $\hat B_{B_q}$, as well as the ratio
$\hat B_{B_s}/\hat B_{B_d}$ from the review~\cite{Battaglia:2003in} as well
as from two more recent evaluations~\cite{Aoki:2003xb,Korner:2003zk} in 
Table~\ref{table1}.
Thus, at least naively, one can determine CKM matrix element $V_{td}$ by measuring $f_B$ 
and $\Delta m_d$ and computing $B_{B_q}$. 

This direct approach, however, meets several difficulties. First,
leptonic decay constant $f_B$ can in principle be extracted from leptonic decays of 
charged B mesons. The corresponding decay width is
\beq
\Gamma(B \to l \nu) = \frac{G_F^2}{8\pi} f_B^2 \left|V_{ub}\right|^2 m_l^2 m_B 
\left(1-\frac{m_l^2}{m_B^2}\right).
\eeq
This width is seen to be quite small due to the smallness of the CKM factor 
$\left|V_{ub}\right|$ and helicity suppression factor of $m_l^2$. In addition,
experimental difficulties are also expected due to the backgrounds stemming from
the presence of a neutrino in the final state. 

\begin{table}
\caption{\label{table1} Renormalization-group independent B-parameters.}
\begin{tabular}{|c|c|c|}
\hline
Method (reference) & $\hat B_{B_d}$ & $\hat B_{B_s}/\hat B_{B_d}$ \\
\hline\hline
Lattice, '03~\cite{Battaglia:2003in} & $1.34(12)$ &
$1.00(3)$ \\
QCDSR, '03~\cite{Battaglia:2003in} & $1.67\pm 0.23$ &
$\approx 1$ \\
Lattice, '03~\cite{Aoki:2003xb} & $1.277(88)(^{+86}_{-95})$ & 
$1.017(16)(^{+56}_{-17})$ \\
QCDSR, '03~\cite{Korner:2003zk} & $1.60\pm 0.03$ & 
$\approx 1$ \\
\hline
\end{tabular}
\end{table}

Second, computation of $B_{B_q}$ is quite difficult and requires the use of non-perturbative 
techniques such as lattice or QCD Sum Rules. Current uncertainties in the determinations
of $f_B$ and $B_{B_q}$ are quite large. It turns out that evaluation of
the ratio of 
\beq
\frac{\Delta m_d}{\Delta m_s} = \frac{m_{B_d}}{m_{B_s}}
\left[\frac{\sqrt{B_{B_d}}f_{B_d}}{\sqrt{B_{B_s}}f_{B_s}}\right]^2
\left|\frac{V_{td}}{V_{ts}}\right|^2,
\eeq
is favored by lattice community, as many systematic errors cancel in this ratio.
This gives a ratio of $\left|V_{td}/V_{ts}\right|$, which provides a non-trivial 
constraint on CKM parameters in the $\rho-\eta$ plane.
 
Instead, one can make use of ample statistics available in charm production
experiments, as heavy quark and $SU(3)$ flavor symmetries relate the ratio  
of charm decay constants $f_{D_s}/f_{D}$ to beauty decay constants 
$f_{B_s}/f_{B}$
\beq
\frac{f_{B_s}/f_{B}}{f_{D_s}/f_{D}}=1+{\cal O}(m_s)\times{\cal O}(1/m_b-1/m_c).
\eeq
Note that SU(3)-violating corrections can also be evaluated in chiral perturbation 
theory~\cite{Grinstein:1993ys}. One still needs to rely on the theoretical 
determination of $B_{B_q}$.

Similar techniques of relating $B$ and $D$ decays can also be used to extract other
CKM matrix elements, like $V_{ub}$~\cite{Ligeti:1997aq}, studies of lifetime 
patterns of heavy hadrons~\cite{Pedrini:2003ee}, and tuning lattice QCD 
calculations~\cite{Juttner:2003cf}.

%%%%%%%%%%%%%%%%%%%%%%%%%%%%%%%%%%%%%%%%%%%
\section{Searching for New Physics}

Another area of modern phenomenology where charm decays play an important 
role is the indirect search for physics beyond the Standard Model. Indeed, large 
statistics usually available in charm physics experiment makes it possible to
probe small effects that might be generated by the presence of
new physics particles and interactions. A program of searches for new physics 
in charm is complimentary to the corresponding programs in bottom or strange 
systems. This is in part due to the fact loop-dominated processes such as
$\DDbar$ mixing or flavor-changing neutral current (FCNC) decays are
sensitive to the dynamics of ultra-heavy {\it down-type particles}. Also, 
in many dynamical models, including the Standard Model, the effects in $s$, $c$, 
and $b$ systems are correlated.

The low energy effect of new physics particles can be 
naturally written in terms of a series of local operators of increasing
dimension generating $\Delta C = 1$ (decays) or $\Delta C = 2$ (mixing) 
transitions. For $\DDbar$ mixing these operators,
as well as the one loop Standard Model effects, generate contributions 
to the effective operators that change $D^0$ state into $\barD$ state
leading to the mass eigenstates
\begin{equation} \label{definition1}
| D_{^1_2} \rangle =
p | D^0 \rangle \pm q | \bar D^0 \rangle,
\end{equation}
where the complex parameters $p$ and $q$ are obtained from diagonalizing 
the $D^0-\barD$ mass matrix. The mass and width splittings between these 
eigenstates are given in Eq.~(\ref{definition}). It is known experimentally 
that $\DDbar$ mixing proceeds extremely slowly, which in the Standard Model 
is usually attributed to the absence of superheavy quarks destroying GIM 
cancellations~\cite{Petrov:1997ch}.

It is instructive to see how new physics can affect charm mixing.
Since the lifetime difference $y$ is constructed from the decays of $D$ into 
physical states, it should be dominated by the Standard Model contributions, 
unless new physics significantly modifies $\Delta C=1$ interactions. On the 
contrary, the mass difference $x$ can receive contributions from all energy scales.
Thus, it is usually conjectured that new physics can significantly
modify $x$ leading to the inequality $x\gg y$~\footnote{This signal for new physics 
is lost if a relatively large $y$, of the order of a percent, 
is observed~\cite{Bergmann:2000id,Falk:2001hx}.}. 

The same considerations apply to FCNC decays as well, where new physics could possibly 
contribute to the decay rates of $D \to X_u \gamma,~D \to X_u l^+ l^-$ 
(with $X_u$ being exclusive or inclusive final state) as well as other 
observables~\cite{FajferProc}. One technical problem here is that
in the standard model these decays are overwhelmingly dominated by long-distance
effects, which makes them extremely difficult to predict model-independently. 
This problem can be turned into a virtue~\cite{Fajfer:2000zx}.

Another possible manifestation of new physics interactions in the charm
system is associated with the observation of (large) CP-violation. This 
is due to the fact that all quarks that build up the hadronic states in weak 
decays of charm mesons belong to the first two generations. Since $2\times2$ 
Cabbibo quark mixing matrix is real, no CP-violation is possible in the
dominant tree-level diagrams that describe the decay amplitudes. 
In the Standard Model CP-violating amplitudes can be introduced by including 
penguin or box operators induced by virtual $b$-quarks. However, their 
contributions are strongly suppressed by the small combination of 
CKM matrix elements $V_{cb}V^*_{ub}$. It is thus widely believed that the 
observation of (large) CP violation in charm decays or mixing would be an 
unambiguous sign for new physics. This fact makes charm decays a valuable 
tool in searching for new physics, since the statistics available in charm 
physics experiment is usually quite large.

As in B-physics, CP-violating contributions in charm can be generally 
classified by three different categories:
(I) CP violation in the decay amplitudes. This type of CP violation 
occurs when the absolute value of the decay amplitude for $D$ to decay to a 
final state $f$ ($A_f$) is different from the one of corresponding 
CP-conjugated 
amplitude (``direct CP-violation'');
(II) CP violation in $\DDbar$ mixing matrix. This type of CP violation
is manifest when 
$R_m^2=\left|p/q\right|^2=(2 M_{12}-i \Gamma_{12})/(2 M_{12}^*-i 
\Gamma_{12}^*) \neq 1$; 
and 
(III) CP violation in the interference of decays with and without mixing.
This type of CP violation is possible for a subset of final states to which
both $D^0$ and $\barD$ can decay. 

For a given final state $f$, CP violating contributions can be summarized 
in the parameter 
\beq
\lambda_f = \frac{q}{p} \frac{{\overline A}_f}{A_f}=
R_m e^{i(\phi+\delta)}\left| \frac{{\overline A}_f}{A_f}\right|,
\eeq
where $A_f$ and ${\overline A}_f$ are the amplitudes for $D^0 \to f$ and 
$\barD \to f$ transitions respectively and $\delta$ is the strong phase 
difference between $A_f$ and ${\overline A}_f$. Here $\phi$ represents the
convention-independent weak phase difference between the ratio of 
decay amplitudes and the mixing matrix.

%%%%%%%%%%%%%%%%%%%%%%%%%%%%%%
\subsection{$\DDbar$ mixing parameters}

Presently, experimental information about the $\DDbar$ mixing parameters 
$x$ and $y$ comes from the time-dependent analyses that can roughly be divided
into two categories. First, more traditional studies look at the time
dependence of $D \to f$ decays, where $f$ is the final state that can be
used to tag the flavor of the decayed meson. The most popular is the
non-leptonic doubly Cabibbo suppressed decay $D^0 \to K^+ \pi^-$.
Time-dependent studies allow one to separate the DCSD from the mixing 
contribution $D^0 \to \D0bar \to K^+ \pi^-$,
\begin{eqnarray}\label{Kpi}
\Gamma[D^0(t) \to K^+ \pi^-]
=e^{-\Gamma t}|A_{K^-\pi^+}|^2 \qquad\qquad\qquad\qquad\qquad
\nonumber \\
\times ~\left[
R+\sqrt{R}R_m(y'\cos\phi-x'\sin\phi)\Gamma t
+\frac{R_m^2}{4}(y^2+x^2)(\Gamma t)^2
\right],
\end{eqnarray}
where $R$ is the ratio of DCS and Cabibbo favored (CF) decay rates. 
Since $x$ and $y$ are small, the best constraint comes from the linear terms 
in $t$ that are also {\it linear} in $x$ and $y$.
A direct extraction of $x$ and $y$ from Eq.~(\ref{Kpi}) is not possible due 
to unknown relative strong phase $\delta_D$ of DCS and CF 
amplitudes~\cite{Falk:1999ts}, 
as $x'=x\cos\delta_D+y\sin\delta_D$, $y'=y\cos\delta_D-x\sin\delta_D$. 
As discussed above, this phase can be 
measured independently~\cite{Silva:1999bd,GGR}. The corresponding formula can 
also be written~\cite{Bergmann:2000id} for $\barD$ decay with $x' \to -x'$ and 
$R_m \to R_m^{-1}$.

Second, $D^0$ mixing can be measured by comparing the lifetimes 
extracted from the analysis of $D$ decays into the CP-even and CP-odd 
final states. This study is also sensitive to a {\it linear} function of 
$y$ via
\beq
\frac{\tau(D \to K^-\pi^+)}{\tau(D \to K^+K^-)}-1=
y \cos \phi - x \sin \phi \left[\frac{R_m^2-1}{2}\right].
\eeq
Time-integrated studies of the semileptonic transitions are sensitive
to the {\it quadratic} form $x^2+y^2$ and at the moment are not 
competitive with the analyses discussed above. 

The construction of new tau-charm factories CLEO-c and 
BES-III will introduce new {\it time-independent} methods that 
are sensitive to a linear function of $y$. One can again use the 
fact that heavy meson pairs produced in the decays of heavy quarkonium 
resonances have the useful property that the two mesons are in the CP-correlated 
states~\cite{AtwoodPetrov} (see Eq.~(\ref{TCFinitial})).

By tagging one of the mesons as a CP eigenstate, a lifetime difference 
may be determined by measuring the leptonic branching ratio of the other meson.
Its semileptonic {\it width} should be independent of the CP quantum number 
since it is flavor specific, yet its {\it branching ratio} will be inversely 
proportional to the total width of that meson. Since we know whether this $D(k_2)$ state is 
tagged as a (CP-eigenstate) $D_\pm$ from the decay of $D(k_1)$ to a 
final state $S_\sigma$ of definite CP-parity $\sigma=\pm$, we can 
easily determine $y$ in terms of the semileptonic branching ratios of $D_\pm$. This 
can be expressed simply by introducing the ratio
\beq \label{DefCor}
R^L_\sigma=
\frac{\Gamma[\psi_L \to (H \to S_\sigma)(H \to X l^\pm \nu )]}{
\Gamma[\psi_L \to (H \to S_\sigma)(H \to X)]~Br(H^0 \to X l \nu)},
\eeq
where $X$ in $H \to X$ stands for an inclusive set of all
final states. A deviation from $R^L_\sigma=1$ implies a
lifetime difference. Keeping only the leading (linear) contributions
due to mixing, $y$ can be extracted from this experimentally obtained 
quantity,
\begin{eqnarray}
y\cos\phi=
(-1)^L {\sigma}
{R^L_\sigma-1\over R^L_\sigma}
\label{y-cos-phi}.
\end{eqnarray}

The current experimental upper bounds on $x$ and $y$ are on the order of 
a few times $10^{-2}$, and are expected to improve significantly in the coming
years.  To regard a future discovery of nonzero $x$ or $y$ as a signal for new 
physics, we would need high confidence that the Standard Model predictions lie
well below the present limits.  As was recently shown~\cite{Falk:2001hx}, 
in the Standard Model, $x$ and $y$ are generated only at second order in SU(3)$_F$ 
breaking, 
\beq
x\,,\, y \sim \sin^2\theta_C \times [SU(3) \mbox{ breaking}]^2\,,
\eeq
where $\theta_C$ is the Cabibbo angle.  Therefore, predicting the
Standard Model values of $x$ and $y$ depends crucially on estimating the 
size of SU(3)$_F$ breaking.  Although $y$ is expected to be determined
by the Standard Model processes, its value nevertheless affects significantly 
the sensitivity to new physics of experimental analyses of $D$ 
mixing~\cite{Bergmann:2000id}.

Theoretical predictions of $x$ and $y$ within and beyond
the Standard Model span several orders of magnitude~\cite{Nelson:1999fg}.
Roughly, there are two approaches, neither of which give very reliable
results because $m_c$ is in some sense intermediate between heavy and
light.  The ``inclusive'' approach is based on the operator
product expansion (OPE).  In the $m_c \gg \Lambda$ limit, where
$\Lambda$ is a scale characteristic of the strong interactions, $\Delta
M$ and $\Delta\Gamma$ can be expanded in terms of matrix elements of local
operators\cite{Inclusive}.  Such calculations yield $x,y < 10^{-3}$.  
The use of the OPE relies on local quark-hadron duality, 
and on $\Lambda/m_c$ being small enough to allow a truncation of the series
after the first few terms.  The charm mass may not be large enough for these 
to be good approximations, especially for nonleptonic $D$ decays.
An observation of $y$ of order $10^{-2}$ could be ascribed to a
breakdown of the OPE or of duality,  but such a large
value of $y$ is certainly not a generic prediction of OPE analyses.
The ``exclusive'' approach sums over intermediate hadronic
states, which may be modeled or fit to experimental data\cite{Exclusive}.
Since there are cancellations between states within a given $SU(3)$
multiplet, one needs to know the contribution of each state with high 
precision. However, the $D$ is not light enough that its decays are dominated
by a few final states.  In the absence of sufficiently precise data on many decay 
rates and on strong phases, one is forced to use some assumptions. While most 
studies find $x,y < 10^{-3}$, Refs.~\cite{Exclusive} obtain $x$ and 
$y$ at the $10^{-2}$ level by arguing that SU(3)$_F$ violation is of order
unity, but the source of the large SU(3)$_F$ breaking is not made explicit.
It was also shown that phase space effects alone provide enough SU(3)$_F$ 
violation to induce $y\sim10^{-2}$~\cite{Falk:2001hx}.
Large effects in $y$ appear for decays close to $D$ threshold, where
an analytic expansion in SU(3)$_F$ violation is no longer possible.
Thus, theoretical calculations of $x$ and $y$ are quite uncertain, and the values
near the current experimental bounds cannot be ruled out. Therefore, it will 
be difficult to find a clear indication of 
physics beyond the Standard Model in $\DDbar$ mixing measurements alone.
The only robust potential signal of new physics in charm system at this stage 
is CP violation.

%%%%%%%%%%%%%%%%%%%%%%%%%%%%%%
\subsection{CP-violation in charm}

CP violation in $D$ decays and mixing can be searched for by a variety of 
methods. For instance, time-dependent decay widths for $D \to K \pi$ are 
sensitive to CP violation in mixing (see Eq.(\ref{Kpi})). Provided that 
the $x$ and $y$ are comparable to experimental sensitivities, a combined 
analysis of $D \to K \pi$ and $D \to KK$ can yield interesting 
constraints on CP-violating parameters~\cite{Bergmann:2000id}.

Most of the techniques that are sensitive to CP violation make use of the
decay asymmetry,
\begin{eqnarray}\label{Acp}
A_{CP}(f)=\frac{\Gamma(D \to f)-\Gamma({\overline D} \to {\overline f})}{
\Gamma(D \to f)+\Gamma({\overline D} \to {\overline f})}=
\frac{1-\left|{\overline A}_{\overline f}/A_f\right|^2}{1+
\left|{\overline A}_{\overline f}/A_f\right|^2}.
\end{eqnarray}
Most of the properties of Eq.~(\ref{Acp}), such as dependence on the
strong final state phases, are similar to the ones in B-physics~\cite{BigiSandaBook}.
Current experimental bounds from various experiments, all consistent
with zero within experimental uncertainties, can be found in~\cite{Pedrini:2000ge}.

Other interesting signals of $CP$-violation that are being 
discussed in connection with tau-charm factory measurements are
the ones that are using quantum coherence of the initial state.
An example of this type of signal is a decay $(D^0 \barD) \to f_1 f_2$ at 
$\psi(3770)$ with $f_1$ and $f_2$ being the different final CP-eigenstates
of the same CP-parity. This type of signals are very easy to detect 
experimentally. The corresponding CP-violating decay rate for the final states
$f_1$ and $f_2$ is
\begin{eqnarray} \label{CPrate}
\Gamma_{f_1 f_2} &=&
\frac{1}{2 R_m^2} \left[
\left(2+x^2-y^2\right) \left|\lambda_{f_1}-\lambda_{f_2}\right|^2
\right.
\nonumber \\
&+& \left .\left(x^2+y^2\right)\left|1-\lambda_{f_1} \lambda_{f_2}\right|^2
\right]~\Gamma_{f_1} \Gamma_{f_2}.
\end{eqnarray}
The result of Eq.~(\ref{CPrate}) represents a generalization of the formula 
given in Ref.~\cite{Bigi:1986dp}. It is clear that both terms in the numerator 
of Eq.~(\ref{CPrate}) receive contributions from CP-violation of the type I 
and III, while the second term is also sensitive to CP-violation of the
type II. Moreover, for a large set of the final states the first term would be 
additionally suppressed by SU(3)$_F$ symmetry, as for instance, 
$\lambda_{\pi\pi}=\lambda_{KK}$ in the SU(3)$_F$ symmetry limit. 
This expression is of the {\it second} order in CP-violating parameters 
(it is easy to see that in the approximation where only CP violation in the mixing 
matrix is retained, $\Gamma_{f_1 f_2} \propto \left|1-R_m^2\right|^2 \propto A_m^2$).
As it follows from the existing experimental 
constraints on rate asymmetries, CP-violating phases are quite small in charm system, regardless 
of whether they are produced by the Standard Model mechanisms or by some new physics 
contributions. In that respect, it looks unlikely that the SM signals of CP violation 
would be observed at CLEO-c with this observable.

While the searches for direct CP violation via the asymmetry of Eq.~(\ref{Acp}) can be
done with the charged D-mesons (which are self-tagging), investigations of the other two 
types of CP-violation require flavor tagging of the initial state. This severely
cuts the available dataset. It is therefore interesting to look for signals of CP violation
that do not require identification of the initial state. One possible CP-violating 
signal involves the observable obtained by summing over the initial 
states, $\sum \Gamma_i=\Gamma_i+{\overline \Gamma}_i$ for $i=f,{\overline f}$.
A CP-odd observable that can be formed out of $\sum \Gamma_i$ is an 
asymmetry~\cite{Petrov:2002qb}
\begin{equation} \label{TotAsym}
A_{CP}^U=\frac{\sum \Gamma_f - \sum \Gamma_{\overline f}}{\sum \Gamma_f + 
\sum \Gamma_{\overline f}}.
\end{equation}
Note that this asymmetry does not require quantum coherence of the initial state 
and therefore is accessible in any D-physics experiment. The final states must be 
chosen such that $A_{CP}^U$ is not trivially zero. It is easy to see that 
decays of $D$ into the final states that are CP-eigenstates 
would result in zero asymmetry, while the final states like $K^+ K^{*-}$ 
or $K_S \pi^+ \pi^-$ would not. A non-zero value of $A_{CP}^U$ in 
Eq.~(\ref{TotAsym}) can be generated by both direct and indirect 
CP-violating contributions. These can be separated by appropriately 
choosing the final states. For example, indirect CP violating amplitudes are
tightly constrained in the decays dominated by the Cabibbo-favored 
tree level amplitudes, while singly Cabibbo suppressed amplitudes 
also receive contributions from direct CP violating amplitudes. 

%%%%%%%%%%%%%%%%%%%%%%%%%%%%%%%%%%%%%%%%%%%
\section{Acknowledgments}

I would like to thank my collaborators as well as the organizers of this workshop 
for making it a success. This work was supported in part by the 
U.S. Department of Energy under grant DE-FG02-96ER41005.

%%%%%%%%%%%%%%%%%%%%%%%%%%%%%%%%%%%%%%%%%%%%%%%%%%%%%%%%%%%%%%


\begin{thebibliography}{9}

\bibitem{Fleischer:2003yc}
R.~Fleischer,
%``Flavour symmetry and clean strategies to extract gamma,''
these proceedings, arXiv:hep-ph/0306270.
%%CITATION = HEP-PH 0306270;%%

\bibitem{Gershon:2003zk}
T.~J.~Gershon,
%``Status and prospects for measurements of phi(3) from B $\to$ D X decays,''
these proceedings, arXiv:hep-ex/0307020.
%%CITATION = HEP-EX 0307020;%%

\bibitem{Gronau:1991dp}
M.~Gronau and D.~Wyler,
%``On determining a weak phase from CP asymmetries in charged B decays,''
Phys.\ Lett.\ B {\bf 265}, 172 (1991).
%%CITATION = PHLTA,B265,172;%%

\bibitem{Atwood:1996ci}
D.~Atwood, I.~Dunietz and A.~Soni,
%``Enhanced CP violation with B $\to$ K D0 (anti-D0) modes and extraction  of the CKM angle gamma,''
Phys.\ Rev.\ Lett.\  {\bf 78}, 3257 (1997).
%%CITATION = HEP-PH 9612433;%%

\bibitem{Grossman:2002aq}
Y.~Grossman, Z.~Ligeti and A.~Soffer,
%``Measuring gamma in B+- $\to$ K+- (K K*)(D) decays,''
Phys.\ Rev.\ D {\bf 67}, 071301 (2003).
%%CITATION = HEP-PH 0210433;%%

\bibitem{Fleischer:2003aj}
R.~Fleischer,
%``A closer look at B/d,s $\to$ D f(r) decays and novel avenues to  determine gamma,''
Nucl.\ Phys.\ B {\bf 659}, 321 (2003).
%%CITATION = HEP-PH 0301256;%%

\bibitem{Giri:2003ty}
A.~Giri, Y.~Grossman, A.~Soffer and J.~Zupan,
%``Determining gamma using B+- $\to$ D K+- with multibody D decays,''
arXiv:hep-ph/0303187;
%%CITATION = HEP-PH 0303187;%%
J.~Zupan, theese proceedings;
D.~Atwood and A.~Soni,
%``Measurement of gamma at B factories using inclusive D decays,''
these proceedings, arXiv:hep-ph/0307154.
%%CITATION = HEP-PH 0307154;%%

\bibitem{Falk:2000ga}
A.~F.~Falk and A.~A.~Petrov,
%``Measuring gamma cleanly with CP-tagged B/s and B/d decays,''
Phys.\ Rev.\ Lett.\  {\bf 85}, 252 (2000);
%%CITATION = HEP-PH 0003321;%%
A.~F.~Falk,
%``CP tagged decays at SuperBaBar,''
Phys.\ Rev.\ D {\bf 64}, 093011 (2001).
%%CITATION = HEP-PH 0107066;%%

\bibitem{Silva:1999bd}
J.~P.~Silva and A.~Soffer,
%``Impact of D0 anti-D0 mixing on the experimental determination of gamma,''
Phys.\ Rev.\ D {\bf 61}, 112001 (2000).
%%CITATION = HEP-PH 9912242;%%

\bibitem{GGR}
M.~Gronau, Y.~Grossman and J.~L.~Rosner,
%``Measuring D0 - anti-D0 mixing and relative strong phases at
%a charm  factory,''
Phys.\ Lett.\ B {\bf 508}, 37 (2001);
%%CITATION = HEP-PH 0103110;%%
E.~Golowich and S.~Pakvasa,
%``Phenomenological issues in the determination of Delta(Gamma(D)),''
Phys.\ Lett.\ B {\bf 505}, 94 (2001).
%%CITATION = HEP-PH 0102068;%%

\bibitem{Rosner:2003yk}
J.~L.~Rosner and D.~A.~Suprun,
%``Measuring the relative strong phase in D0 $\to$ K*+ K- and D0 $\to$ K*- K+  decays,''
arXiv:hep-ph/0303117.
%%CITATION = HEP-PH 0303117;%%

\bibitem{Bigi:1999hr}
I.~I.~Bigi and A.~I.~Sanda,
%``On the other five KM triangles,''
arXiv:hep-ph/9909479.
%%CITATION = HEP-PH 9909479;%%

\bibitem{Abele:2002wc}
H.~Abele {\it et al.},
%``Is the unitarity of the quark-mixing-CKM-matrix violated in neutron  beta-decay?,''
Phys.\ Rev.\ Lett.\  {\bf 88}, 211801 (2002).
%%CITATION = HEP-EX 0206058;%%

\bibitem{Buchalla:1995vs}
G.~Buchalla, A.~J.~Buras and M.~E.~Lautenbacher,
%``Weak Decays Beyond Leading Logarithms,''
Rev.\ Mod.\ Phys.\  {\bf 68}, 1125 (1996);
%%CITATION = HEP-PH 9512380;%%
A.~J.~Buras, M.~Jamin and P.~H.~Weisz,
%``Leading And Next-To-Leading QCD Corrections To Epsilon 
% Parameter And B0 - Anti-B0 Mixing In The Presence Of A Heavy Top Quark,''
Nucl.\ Phys.\ B {\bf 347}, 491 (1990);
%%CITATION = NUPHA,B347,491;%%
M.~Ciuchini, E.~Franco, G.~Martinelli, L.~Reina and L.~Silvestrini,
%``An Upgraded analysis of epsilon-prime epsilon at the next-to-leading order,''
Z.\ Phys.\ C {\bf 68}, 239 (1995).
%%CITATION = HEP-PH 9501265;%%

% B_{B} references

\bibitem{Battaglia:2003in}
M.~Battaglia {\it et al.},
%``The CKM matrix and the unitarity triangle,''
arXiv:hep-ph/0304132.
%%CITATION = HEP-PH 0304132;%%

\bibitem{Aoki:2003xb}
S.~Aoki {\it et al.}  [JLQCD Collaboration],
%``B0 anti-B0 mixing in unquenched lattice QCD,''
arXiv:hep-ph/0307039.
%%CITATION = HEP-PH 0307039;%%

\bibitem{Korner:2003zk}
J.~G.~Korner, A.~I.~Onishchenko, A.~A.~Petrov and A.~A.~Pivovarov,
%``B0 anti-b0 mixing beyond factorization,''
arXiv:hep-ph/0306032.
%%CITATION = HEP-PH 0306032;%%

% end of B_{B} references

\bibitem{Grinstein:1993ys}
B.~Grinstein,
%``On a precise calculation of (f(B(s)) / f(B)) / (f(D(s)) / f(D)) 
% and its implications on the interpretation of B anti-B mixing,''
Phys.\ Rev.\ Lett.\  {\bf 71}, 3067 (1993).
%%CITATION = HEP-PH 9308226;%%

\bibitem{Ligeti:1997aq}
Z.~Ligeti, I.~W.~Stewart and M.~B.~Wise,
%``Comment on V(ub) from exclusive semileptonic B and D decays,''
Phys.\ Lett.\ B {\bf 420}, 359 (1998).
%%CITATION = HEP-PH 9711248;%%

\bibitem{Pedrini:2003ee}
D.~Pedrini  [the FOCUS Collaboration],
%``Recent results on charm from E831-FOCUS,''
these proceedings, arXiv:hep-ph/0307137;
%%CITATION = HEP-PH 0307137;%%
M.~B.~Voloshin,
%``Relations between inclusive decay rates of heavy baryons,''
Phys.\ Rept.\  {\bf 320}, 275 (1999);
%%CITATION = HEP-PH 9901445;%%
B.~Guberina, B.~Melic and H.~Stefancic,
%``Lifetime-difference pattern of heavy hadrons,''
Phys.\ Lett.\ B {\bf 484}, 43 (2000).
%%CITATION = HEP-PH 0004264;%%

\bibitem{Juttner:2003cf}
P.~Mackenzie, these proceedings;
A.~Juttner and J.~Rolf,
%``Precision computation of the leptonic D/s-meson decay constant in  quenched QCD,''
these proceedings, arXiv:hep-ph/0306299;
%%CITATION = HEP-PH 0306299;%%

\bibitem{Petrov:1997ch}
A.~Datta, D.~Kumbhakar,
%``D0 Anti-D0 Mixing: A Possible Test Of Physics Beyond The Standard Model,''
Z.\ Phys.\ C{\bf 27}, 515 (1985);
%%CITATION = ZEPYA,C27,515;%%
A.~A.~Petrov,
%``On dipenguin contribution to D0 anti-D0 mixing,''
Phys.\ Rev.\ D{\bf 56}, 1685 (1997);
%%CITATION = HEP-PH 9703335;%%
E.~Golowich and A.~A.~Petrov,
%``Can nearby resonances enhance D0 anti-D0 mixing?,''
Phys.\ Lett.\ B {\bf 427}, 172 (1998).
%%CITATION = HEP-PH 9802291;%%

\bibitem{Bergmann:2000id}
S.~Bergmann, Y.~Grossman, Z.~Ligeti, Y.~Nir, A.~Petrov,
%``Lessons from CLEO and FOCUS measurements of D0 anti-D0 mixing  parameters,''
Phys.\ Lett.\ B {\bf 486}, 418 (2000).
%%CITATION = HEP-PH 0005181;%%

\bibitem{Falk:2001hx}
A.~F.~Falk, Y.~Grossman, Z.~Ligeti and A.~A.~Petrov,
%``SU(3) breaking and D0 - anti-D0 mixing,''
Phys.\ Rev.\ D {\bf 65}, 054034 (2002).
%%CITATION = HEP-PH 0110317;%%

\bibitem{FajferProc}
S.~Fajfer,
%``Rare decays of D mesons,''
these proceedings, arXiv:hep-ph/0306263.
%%CITATION = HEP-PH 0306263;%%

\bibitem{Fajfer:2000zx}
S.~Fajfer, S.~Prelovsek, P.~Singer and D.~Wyler,
%``A possible arena for searching new physics: 
% The Gamma(D0 $\to$ rho0  gamma)/Gamma(D0 $\to$ omega gamma) ratio,''
Phys.\ Lett.\ B {\bf 487}, 81 (2000).
%%CITATION = HEP-PH 0006054;%%

\bibitem{Falk:1999ts}
A.~F.~Falk, Y.~Nir and A.~A.~Petrov,
%``Strong phases and D0 anti-D0 mixing parameters,''
JHEP {\bf 9912}, 019 (1999).
%%CITATION = HEP-PH 9911369;%%

\bibitem{AtwoodPetrov}
D.~Atwood and A.~A.~Petrov,
%``Lifetime Differences in Heavy Mesons With Time Independent Measurements,''
arXiv:hep-ph/0207165.
%%CITATION = HEP-PH 0207165;%%

\bibitem{Nelson:1999fg}
H.~N.~Nelson,
%``Compilation of D0 $\to$ anti-D0 mixing predictions,''
in {\it Proc. of the 19th Intl. Symp. on Photon and Lepton 
Interactions at High Energy LP99 } ed. J.A. Jaros and M.E. Peskin,
arXiv:hep-ex/9908021.
%%CITATION = HEP-EX 9908021;%%

\bibitem{Inclusive}
H.~Georgi, Phys. Lett. B297, 353 (1992);
%%CITATION = HEP-PH 9209291;%%
T.~Ohl, G.~Ricciardi and E.~Simmons, Nucl. Phys. B403, 605 (1993);
%%CITATION = HEP-PH 9301212;%%
I.~Bigi and N.~Uraltsev,
Nucl.\ Phys.\ B {\bf 592}, 92 (2001),
%%CITATION = HEP-PH 0005089;%%
for a recent review see
A.~A.~Petrov, {\it Proc. of 4th Workshop on Continuous 
Advances in QCD}, Minneapolis, Minnesota, 12-14 May 2000, 
arXiv:hep-ph/0009160.
%%CITATION = HEP-PH 0009160;%%

\bibitem{Exclusive}
J. Donoghue, E. Golowich, B. Holstein and J. Trampetic,
Phys. Rev. D33, 179 (1986);
%%CITATION = PHRVA,D33,179;%%
L. Wolfenstein, Phys.\ Lett.\ B164, 170 (1985);
%%CITATION = PHLTA,B164,170;%%
P. Colangelo, G. Nardulli and N. Paver,  Phys.\ Lett.\ B242, 71 (1990);
%%CITATION = PHLTA,B242,71;%%
T.A. Kaeding,  Phys. Lett. B357, 151 (1995).
%%CITATION = HEP-PH 9505393;%%
A.~A.~Anselm and Y.~I.~Azimov,
%``CP Violating Effects In E+ E- Annihilation,''
Phys.\ Lett.\ B {\bf 85}, 72 (1979);
%%CITATION = PHLTA,B85,72;%%

\bibitem{BigiSandaBook}
I.~I.~Bigi and A.~I.~Sanda,
{\it CP violation} (Cambridge University Press, 2000).

\bibitem{Pedrini:2000ge}
D.~Pedrini,
%``Mixing and {CP} violation in the charm sector,''
J.\ Phys.\ G {\bf 27}, 1259 (2001).
%%CITATION = JPHGB,G27,1259;%%

\bibitem{Bigi:1986dp}
I.~I.~Bigi and A.~I.~Sanda,
%``On D0 - Anti-D0 Mixing And CP Violation,''
Phys.\ Lett.\ B {\bf 171}, 320 (1986).
%%CITATION = PHLTA,B171,320;%%

\bibitem{Petrov:2002qb}
A.~A.~Petrov,
%``CP-violation and mixing in charmed mesons,''
Proc. of the {\it 5th Workshop on Continuous Advances in QCD}, 
pp. 102-114; arXiv:hep-ph/0209049.
%%CITATION = HEP-PH 0209049;%%

\end{thebibliography}
\end{document}